\documentclass[prb,aps,onecolumn,showpacs,floats,a4paper,times,12pt]{revtex4}
\usepackage{epsfig,bm,epsf,graphics}
\begin{document}
\draft
\title{Extended Defects in the Potts-Percolation Model of a Solid: Renormalization Group and Monte Carlo Analysis}
\author{H. T. Diep$^a$\footnote{ Corresponding author, E-mail:diep@u-cergy.fr }  and Miron Kaufman$^{b}$}
\address{
$^a$ Laboratoire de Physique Th\'eorique et Mod\'elisation,
Universit\'e de Cergy-Pontoise, CNRS, UMR 8089\\
2, Avenue Adolphe Chauvin, 95302 Cergy-Pontoise Cedex, France\\
$^b$ Department of Physics, Cleveland State University, Cleveland, OH.44115, USA\\}

\begin{abstract}

We extend the model of a 2$d$ solid to include a line of defects. Neighboring atoms on the defect line are connected by "springs" of different strength and different cohesive energy with respect to the rest of the system. Using the Migdal-Kadanoff renormalization group we show that the elastic energy is an irrelevant field at the bulk critical point. For zero elastic energy this model reduces to the Potts model. By using Monte Carlo simulations of the 3- and 4-state Potts model on a square lattice with a line of defects, we confirm the renormalization-group prediction that for a defect interaction larger than the bulk interaction the order parameter of the defect  line changes discontinuously while the defect energy varies continuously as a function of temperature at the bulk critical temperature.

\end{abstract}

\maketitle
\section{Introduction}
Thermodynamics of solids with defects is a topic of current interest.\cite{Gomez,Man}   In this paper we expand an equilibrium statistical mechanics model\cite{Blumberg} of a solid to include extended defects.  Previously we used\cite{Kaufman96} the realistic anharmonic energy versus atomic distance developed by Ferrante and collaborators\cite{Ferrante} in a mean-field type computation.  We then evaluated\cite{Kaufman2008} the role of thermal fluctuations by using renormalization-group and Monte-Carlo simulations.

In our model described in Section II, the solid is constituted of harmonic springs.  If the energy of such a spring is larger than a threshold, the spring is likely to fail.\cite{Hassold}  In the limit of zero elastic energy the model reduces to the Potts model,\cite{Potts} which has been used to describe correlated and uncorrelated percolation processes.\cite{Wu,Kaufman84b,Kaufman84c,Ivanskoi}

	In Section III we present phase diagrams based on the renormalization-group Migdal-Kadanoff scheme.  This scheme is of course an approximation for models on regular lattices but as demonstrated by Berker\cite{Berker} it is exact for hierarchical lattices.  This latter feature and its simplicity make the Migdal-Kadanoff scheme quite popular.\cite{Griffiths82,Kaufman84d,Kaufman84,Erbas,Guven}  We analyze the stability of fixed points and find that the elastic energy is an irrelevant field in the renormalization-group sense. For this reason, in Section IV we report Monte-Carlo simulations of the model with zero elastic energy, i. e. Potts model on a square lattice with a defect line.  Our goal is to verify the renormalization group prediction\cite{Kaufman82} that for $q = 3, 4$, on the defect line, at the bulk critical temperature, the order parameter jumps discontinuously from zero (high temperature phase) to a nonzero value (low temperature phase)while the energy varies continuously with temperature.  This is interesting since for those q values the bulk transition is continuous.  It was argued\cite{Kaufman82} that this unusual $1d$ defect transition is due to the infinite range correlations at the bulk critical point.  It resembles the Thouless transition\cite{Thouless,Aizenman} in a one-dimensional system with inverse squared distance decaying interactions.
Our concluding remarks are found in Section V.

\section{Model }

The solid is made of "springs" some of which are live and some are failed upon thermal excitations.  All processes are assumed to be reversible, unlike the work of Beale and Srolovitz\cite{Beale} where springs fail irreversibly.  The energy of a "spring" $<i,j>$ is given by the Hooke law:
\begin{equation}
H_{ij}=-E_C+\frac{k}{2}(\vec r_i-\vec r_j)^2
\end{equation}
where $\vec r_i$ is the displacement vector from the equilibrium position of atom $i$,  $E_C$ is the cohesive energy, $k$ is elastic constant and $a$ is the equilibrium lattice spacing.  If the energy of the spring is larger than the threshold energy $E_0$ the "spring" is more likely to fail than to be alive. $p$ is the probability that the "spring" is alive and $1- p$ the probability that the "spring" breaks.  We assume its dependence on energy to be given by the Boltzman weight:

\begin{equation}
\frac{p}{1-p}=e^{-\frac{{\cal
H}-E_0}{k_BT}}=we^{-\frac{K}{2}(\vec r_i-\vec r_j)^2}
\end{equation}

where $K= k/k_BT$ and $w=e^{\frac{E_C+E_0}{k_BT}}$.

For the extended line of defects the elastic, cohesive and threshold energies may take values different from the rest of the system. Hence while in the bulk the parameters are $K$ and $w$, on the line of defects they are $K_d$ and $w_d$.

We allow for correlations between failing events by using the Potts number of states $q$, which plays the role of a fugacity controlling the number of clusters.  For $q = 1$ we have random percolation as springs fail independently.  The partition function is a sum over all possible configurations of "live" springs:

\begin{equation}
Z=\sum_{config}q^cw^B Z_{\mbox{elastic}}^{\mbox{config}}
\end{equation}
 $C$ is the number of clusters, including single site clusters, and $B$ is number of live "springs". The restricted partition function associated with the elastic energy for a given configuration of bonds (live "springs") is

\begin{equation}
Z_{\mbox{elastic}}^{\mbox{config}}=Tr_r e^{
-\frac{H_{\mbox{elastic}}}{k_BT}}
\end{equation}

\begin{equation}\label{H1}
-\frac{H_{\mbox{elastic}}}{k_BT}=\sum_{<i,j>}\frac{K}{2}(\vec r_i-\vec r_j)^2
\end{equation}

In Eq. (\ref{H1}) the sum is over all live "springs".

 By using the Kasteleyn-Fortuin expansion\cite{Fortuin} for Potts model we can rewrite the partition function as

\begin{equation}
Z=Tr_\sigma Tr_r e^{-\frac{{\cal H}}{k_BT}}
\end{equation}

The Hamiltonian is

\begin{equation}\label{H}
-\frac{{\cal H}}{k_BT}=\sum_{<i,j>} [J_1 \delta(\sigma_i,\sigma_j)
-\frac{J_2}{2}\delta(\sigma_i,\sigma_j) (\vec r_i-\vec r_j)^2]+
\sum_{<i,j>defect}[J_{1d} \delta(\sigma_i,\sigma_j)
-\frac{J_{2d}}{2}\delta(\sigma_i,\sigma_j)(\vec r_i-\vec r_j)^2]
\end{equation}
where $\sigma_i$ is a Potts spin taking $q$ values.  This mapping
is a Gaussian approximation valid when, on the right hand side of
Eq. \ref{H}, the elastic energy is small compared to the first
energy contribution.  In our Monte Carlo simulations, we use the Hamiltonian in Eq. (7) for integer values of q.  The coupling constants $J_1$ and $J_2$ are
related to the original parameters, $w$ and $K$, as follows:

\begin{eqnarray}
J_1&=&\ln (1+w)\label{J1}\\
J_2&=&K\frac{w}{w+1}\label{J2}\\
J_{1d}&=&\ln (1+w_d)\label{J1d}\\
J_{2d}&=&K_d
\frac{w_d}{w_d
+1}\label{J2d}
\end{eqnarray}

\section{ Renormalization Group}
The Migdal-Kadanoff recursion equations\cite{Migdal,Kadanoff} for two dimensions are obtained by assuming that each atom coordinate varies in the interval (-1/2, 1/2), where the equilibrium lattice constant is 1, and also using the Gaussian approximation (small elastic energy):

\begin{eqnarray}
w'&=&[1+U(w,K,q)]^2-1\label{WK}\\
K'w'&=&K[1+U(w,K,q)]U(w,K,q)\label{WK1}\\
w_d'&=&[1+U(w,K,q)][1+U(w_d,K_d,q)]-1\label{WK2}\\
K_d'w_d'&=&\frac{1}{2}K_d[1+U(w,K,q)]U(w_d,K_d,q)+
\frac{1}{2}K[1+U(w_d,K_d,q)]U(w,K,q)\label{WK3}
\end{eqnarray}							

where

\begin{equation}
U(w,K,q)=\frac{w^2
erf(\sqrt{K/4})}{q\sqrt{K/\pi}+\sqrt{8}w.erf(\sqrt{K/8})}
\end{equation}

The above recursion equations represent the Gaussian approximation of the exact solutions for hierarchical lattices.
The renormalization group flows in the bulk parameter space $(w, K)$ are governed by the following fixed points at $K = 0$ (pure Potts model):

i. $w = 0$ (non-percolating live "springs"),

ii. $w=\infty$ (percolating network of live "springs"),

iii. $w= w_c$ (Potts critical point).

A stability analysis at the bulk Potts critical point, ($K = 0$, $w = w_c$) yields the two eigenvalues:

i. the thermal eigenvalue: $\Lambda_1$ (for the direction along the $K= 0$ axis) is always larger than 1, meaning the $w- w_c$ is a relevant field;

ii. The other eigenvalue associated with the flow along the $w = w_c$ line away from the pure model ($K = $0) is $\Lambda_2< 1$  for all $q$.  This means that there is a line of points in the $(w, K)$ flowing into, and thus is in the same universality class as, the pure Potts critical point ($w_c, 0$).

\begin{figure}[t!]
\centering
\includegraphics[width=5.0in]{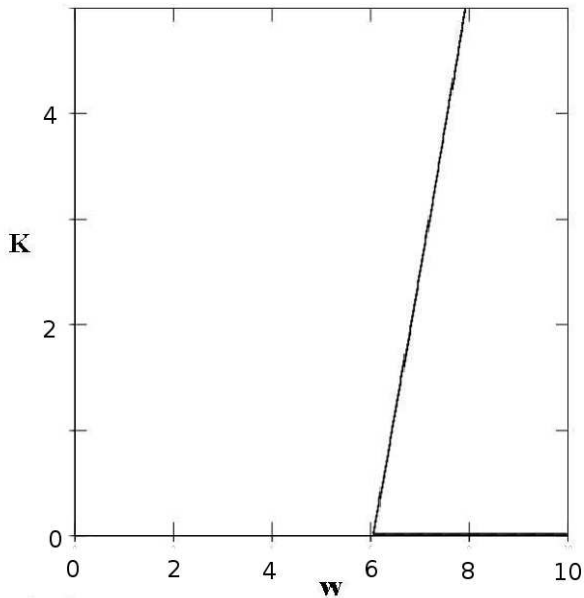}
\caption{Bulk phase diagram for $q$ = 10 and $d$ = 2, in the plane $(w, K)$.} \label{fig:1}
\end{figure}

The bulk phase diagram (Fig.\ref{fig:1}), for any given $q$, in the ($w, K$) plane shows two phases:

I. solid with mostly live springs,

II. "crumbling" solid with failed springs, separated by a critical line in the universality class of the $q$-state Potts model.

The defect fixed points are obtained from  Eqs. (\ref{WK2}) and (\ref{WK3}) after setting the bulk fields $w$ and $K$ at their fixed point values (bulk material is critical): $w = w_c$, $K = 0$.  All the fixed points are obtained for $K_d = 0$.  We find fixed points at $w_d =\infty$ and at $ w_d = w_c$, for all $q$.  For $q > q_1 = 3$ a third fixed point emerges from the one at $w_d =\infty$ at $w_d = w*$. At $q = q_2 = 6.8$ (when exponent $\alpha = 0$) the two finite $w_d$ fixed points exchange position.

\begin{figure}[t!]
\centering
\includegraphics[width=5.0in]{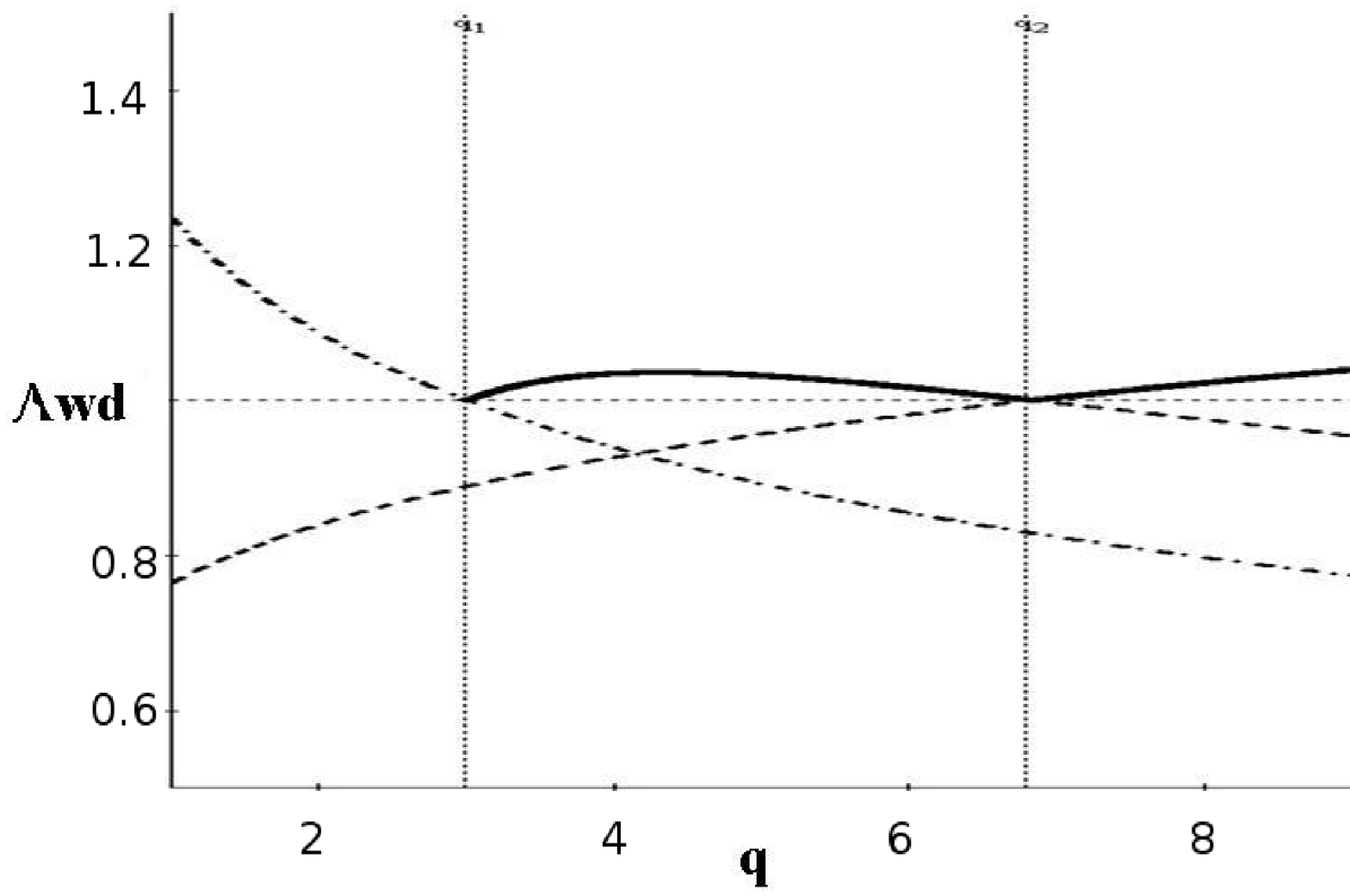}
\caption{Renormalization group eigenvalue $\Lambda w_d$ vs $q$: solid line at defect critical fixed point; dotted-dashed  line at the non-percolating critical fixed point; dashed line at the percolating critical fixed point.  Two vertical lines are $q_1=3$ and $q_2=6.8$ (see text for comments).} \label{fig:2}
\end{figure}

\begin{figure}[t!]
\centering
\includegraphics[width=5.0in]{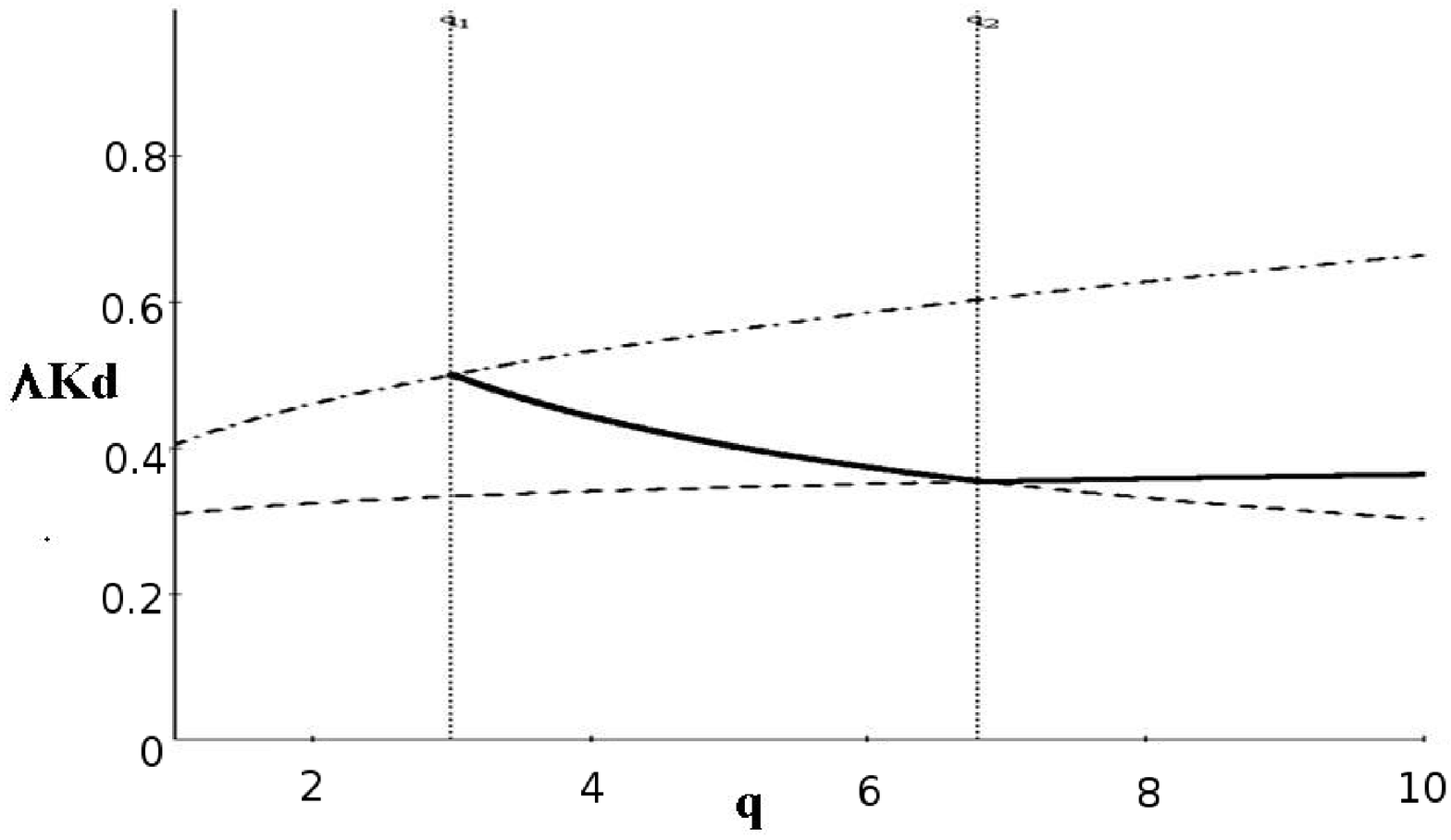}
\caption{Renormalization group eigenvalue $\Lambda K_d$ vs $q$: solid  line at defect critical fixed point; dashed-dotted line at the non-percolating critical fixed point; dashed line at the percolating critical fixed point.  Since these eigenvalues are less than unity, the elastic energy is an irrelevant field at the three fixed points.   Two vertical lines are $q_1=3$ and $q_2=6.8$ (see text for comments).} \label{fig:3}
\end{figure}

The stability of the fixed points for perturbations of the defect fields $(w_d, K_d)$ is determined by

i. The eigenvalue $\Lambda w_d$ governing flow along $K_d = 0$ axis is larger than unity (i.e. relevant perturbation) at the fixed point $ K_d= 0$, $w_d = max(w_c,w*)$ and is less than unity (i.e. irrelevant perturbation) at the fixed point $K_d = 0$, $w_d = min(w_c,w*)$ (Fig. \ref{fig:2});

ii. the eigenvalue $\Lambda K_d  < 1$ for all fixed points (Fig. \ref{fig:3}).

\begin{figure}[t!]
\centering
\includegraphics[width=5.0in]{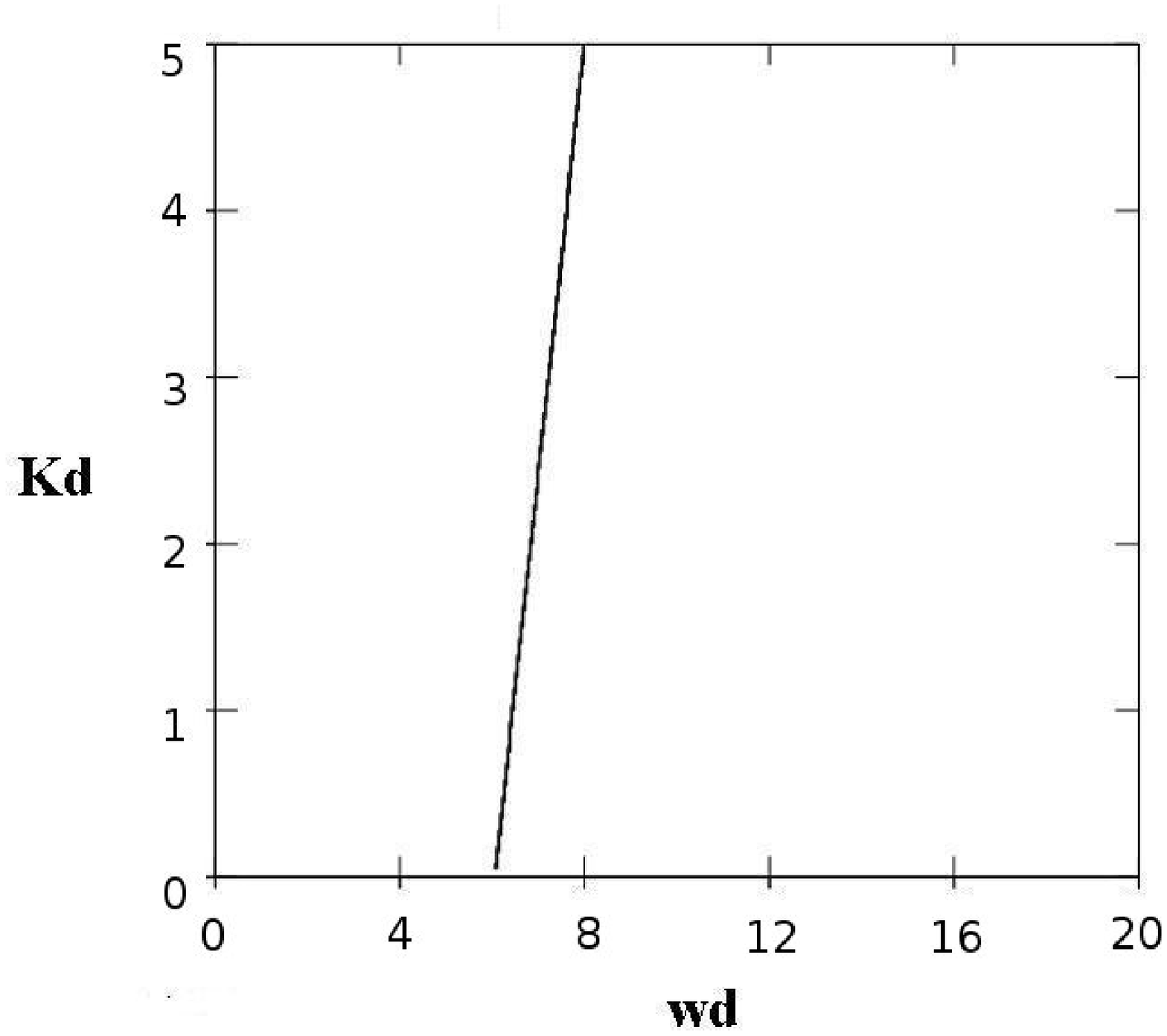}
\caption{Phase diagram at $w = w_c$, $K = 0$ in plane $(w_d, K_d)$. The two critical phases (infinite correlation length) are separated by a defect critical line.   In the small $w_d$ phase there is zero probability for percolation on the defect line, while in the large $w_d$ phase there is a finite probability for percolation on the defect line.} \label{fig:4}
\end{figure}

The defect phase diagram (Fig. \ref{fig:4}) represented in the plane $(w_d, K_d)$ is obtained for the bulk fields at their critical values: $w = w_c$, $K = 0$.

The plane $(w_d, K_d)$ is divided in two regions by a critical line that flows into the fixed point: $w_d = max(w_c,w*)$ , $K_d = 0$. To the right of it (large $w_d$), the live springs on the defect line are percolating, i.e. the Potts order parameter is nonzero on the defect line even though in the bulk it is zero.  All points in that region flow to the fixed point $w_d = \infty$, $K_d = 0$.  The phase to the left of the critical line (small $w_d$) is governed by the fixed point at $w_d = min(w_c,w*)$, $K_d = 0$.  This is a Berezinskii-Kosterlitz-Thouless
\cite{Berezinskii,Kosterlitz,Hinczewski} phase characterized by infinite correlation length or algebraic decay of correlations along the defect line and zero Potts order parameter.

Since the elastic energy is irrelevant RG field, we concentrate next on the $K = K_d = 0$ case. The RG analysis\cite{Kaufman82} of this problem (two-dimensional Potts model with a defect line) predicts for $2 < q \leq4$ a hybrid phase transition on the defect line: discontinuous order parameter and continuous energy.

\section{Monte Carlo Simulations}

In this section since we set the elastic couplings to zero in Eq. (\ref{H}), we perform MC simulation of the
Hamiltonian
\begin{equation}\label{HMC}
{-\frac{{\cal H}}{k_BT}}= J\sum_{<i,j>} \delta(\sigma_i,\sigma_j) +
J_d\sum_{<i,j>defect}\delta(\sigma_i,\sigma_j)
\end{equation}
where $J$ is the interaction parameter between the
bulk nearest-neighbors (NN) and $J_d$ is that between two NN on the defect line ($J$ is equal to $J_1$ of Eq. \ref{J1}, and $J_d$ is $J_{1d}$ of Eq. \ref{J1d}).

We consider a square lattice of
size $N_x\times N_y$ where $N_x=80,120,160,200,...,1920$ and $N_y=40,60,80$.  Each lattice site is
occupied by a $q-$state Potts spin.  We place the defect line at the middle of the
$y$ side of the sample, i.e. $y=N_y/2$.  The length of the defect line is thus $N_x$. The reason why we use a very long length for the defect line
stems from the fact that the determination of a transition in a linear chain needs large enough sizes to avoid statistical fluctuations.  We use periodic boundary
conditions.

 Our purpose here is to test the following RG prediction of the previous section: at the bulk transition temperature $T_c$, the defect line undergoes a phase transition  where its order parameter
is discontinuous but its energy is continuous.   The bulk transition temperature is the temperature at which the phase transition of the system without the defect line takes place.

Let us consider the case where $q=4$.  The critical temperature is given by the exact formula
$(k_BT_c/J)^{-1}=\ln (1+\sqrt{q})$. With $q=4$, one has $T_c \simeq 0.910239$ in units of $J/k_B$.
Note that this value of $T_c$ corresponds to the thermodynamic limit, i.e. infinite system size.
In MC simulation, we work at finite sizes, so for each size we have to determine the "pseudo" transition which corresponds in general to the maximum of the specific heat or of the susceptibility. The
maxima of these quantities need not to be at the same temperature. Only at the infinite size, they should coincide.  The theory of finite-size scaling permits to deduce properties of a system at its thermodynamic limit.  We have used in this work a size large enough to reproduce the bulk transition temperature up to the fourth decimal.

In order to determine the nature of the phase transition of the defect line, we shall use the histogram technique\cite{Ferren} which is
known to allow us to distinguish with accuracy
the order of the phase transition.

The simulation is carried out as follows.   We fix $J=1$ hereafter.  For each value of
$J_d$, using first the standard Metropolis MC method\cite{Binder}
 we equilibrate the system of a given size $N_x\times N_y$ at a given temperature $T$
during $10^6$ Monte Carlo sweeps (MCS) per spin before averaging
physical quantities over the next $2\times 10^6$ MCS.   We determine the transition
temperature at the given size $N_x\times N_y$ by examining the calculated physical quantities such as the
internal energy per spin $E$, the specific heat $C_v$ per spin,
the Potts order parameter $Q$ and the susceptibility per spin
$\chi$. For the bulk $q$-state Potts model,  $Q$ is defined as

\begin{equation}
Q=\frac{q \max (Q_1,Q_2,..., Q_q)-1}{q-1} \nonumber
\end{equation}
where $Q_i=\frac{n_i}{N_x\times N_y}$ ($i=1,...,q)$, $n_i$ being the number
of sites having $q_i$.  For the defect line, the order parameter $Q_d$ is similarly defined
on the defect line, namely

\begin{equation}
Q_d=\frac{q \max (Q'_1,Q'_2,..., Q'_q)-1}{q-1} \nonumber
\end{equation}
where $Q'_i=\frac{n'_i}{N_x}$ ($i=1,...,q)$, $n'_i$ being the number
of sites on the defect line having $q_i$.

\begin{figure}
\centerline{\epsfig{file=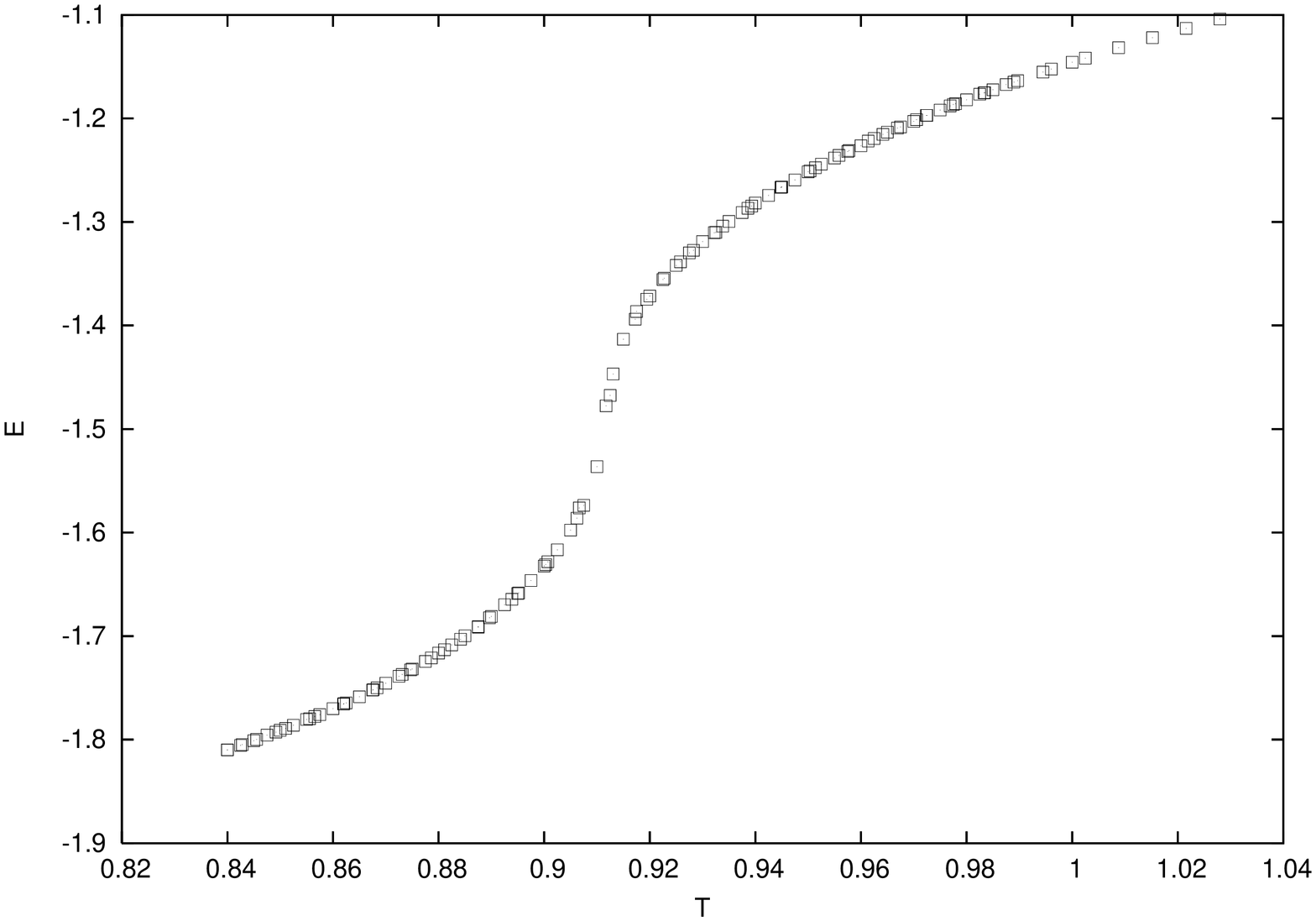,width=2.in,angle=-90}}
\centerline{\epsfig{file=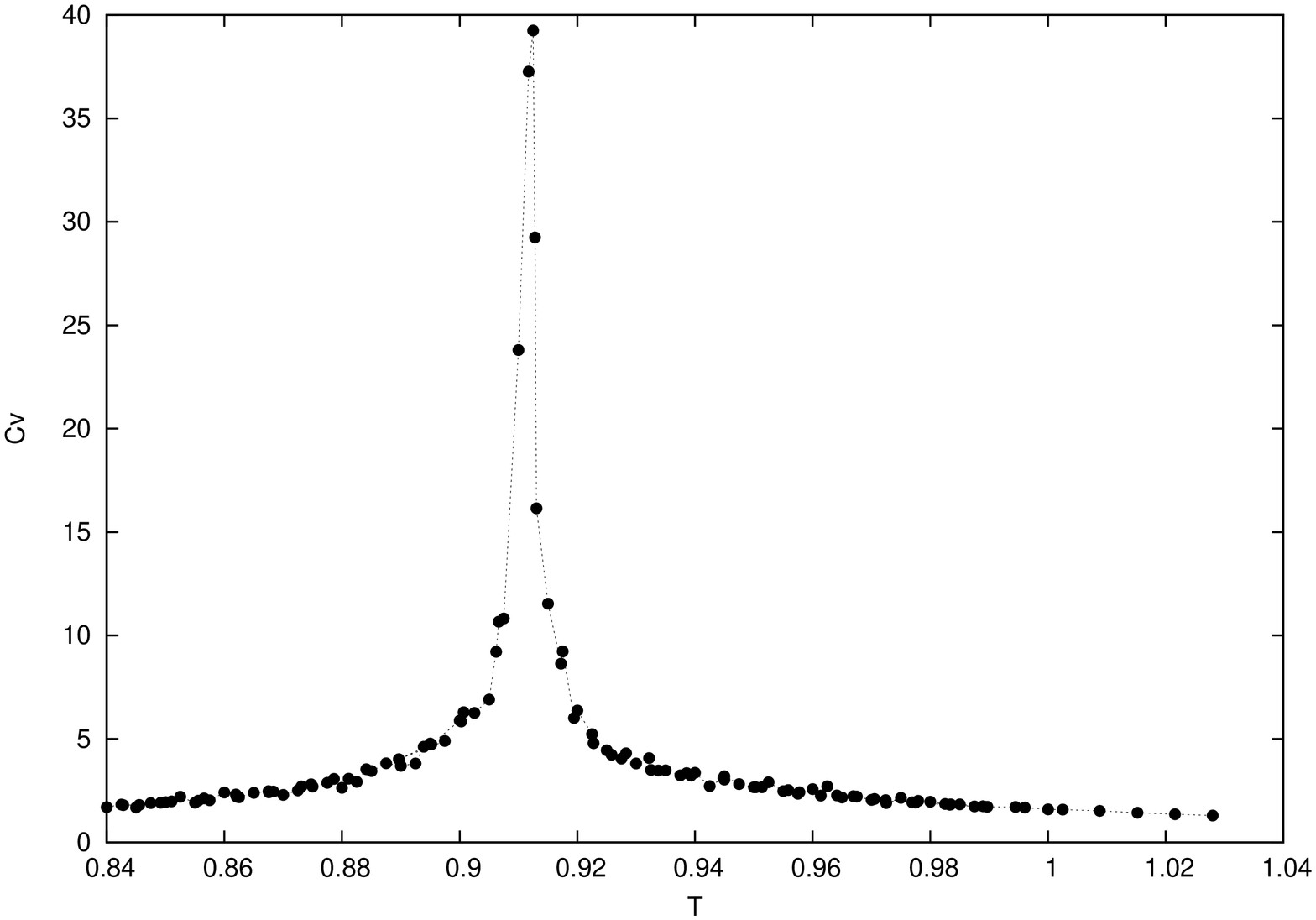,width=2.in,angle=-90}}
\caption{Case without defects: Energy per spin $E$ (upper curve) and specific heat per
spin $C_v$ (lower curve, line is a guide to the eye) vs temperature $T$ for $q=4$,
with $N_x=1920$, $N_y=60$ and $J_d =J = 1$. } \label{fig:E}
\end{figure}

Let us show first in Fig. \ref{fig:E} the
energy and the specific heat of the case without defects where the bulk critical temperature is
$T_c\simeq 0.9103$ for $q=4$ with the size used here (the exact value of $T_c$ is 0.910239).  The energy of the defect line is shown in Fig. \label{fig:ED}

\begin{figure}
\centerline{\epsfig{file=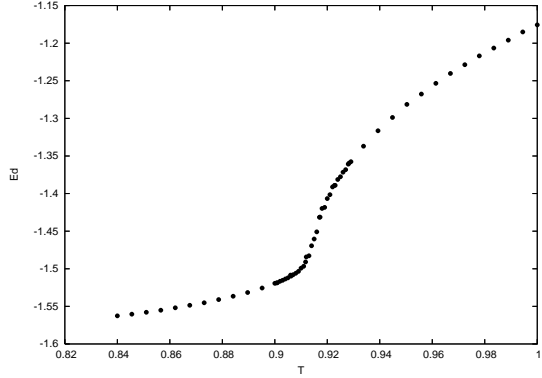,width=2.in,angle=-90}}
\caption{Energy per spin $E_d$ of the defect line vs temperature $T$ for $q=4$, $J_d =1.6$
with $N_x=1920$, $N_y=60$ and $J = 1$. Note that $E_d$ is continuous at $T_c$. See text for comments.} \label{fig:ED}
\end{figure}

\begin{figure}
\centerline{\epsfig{file=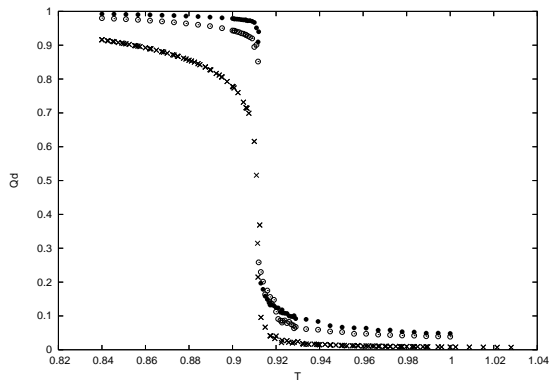,width=2.in,angle=-90}}
\caption{Order parameters  $Q_d$ of the defect line for $J_d = 2$ (upper curve with black circles), $J_d = 1.6$ (middle curve with void circles) and bulk order parameter $Q$ (lower curve with crosses)
vs temperature $T$ for $q=4$
with $N_x=1920$, $N_y=60$ and $J =1$. Note that $Q_d$ is discontinuous at $T_c$. See text for comments.} \label{fig:Q}
\end{figure}

The order parameters the bulk and of the defect line are shown in Fig. \ref{fig:Q} for $q=4$, $J=1$ and $J_d=1.6$ and 2.

These figures show that at the "bulk" phase transition temperature $T_c=0.9103$, while the defect energy
is continuous,
the order parameter of the defect line undergoes a  vertical fall, indicating a discontinuity predicted by the RG analysis shown in the previous section. For large values of $J_d$, for instance $J_d=2$, the defect line takes a very long time to become disordered for $T$ slightly larger than $T_c$.  Several millions MC sweeps are necessary.

\begin{figure}
\centerline{\epsfig{file=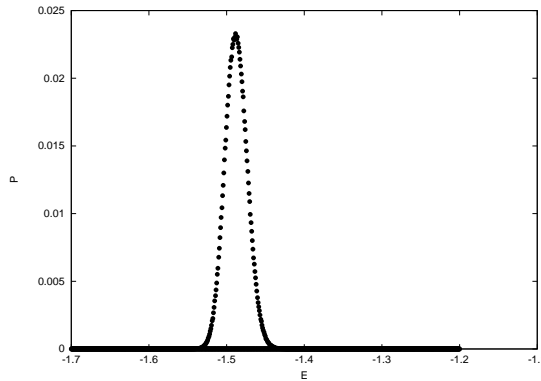,width=2.in,angle=-90}}
\caption{Energy histogram of the defect line taken at $T_c$, for $q=4$, $J_d =1.6$
with $N_x=1920$, $N_y=60$ and $J = 1$.  See text for comments.} \label{fig:HED}
\end{figure}

In order to check the behaviors of $E_d$ and $Q_d$ at $T_c$, we have calculated the defect energy histogram not only at $T_c$ but also
in the temperature region around $T_c$.  As it turned out, we observe only a gaussian distribution of $E_d$
(see Fig. \ref{fig:HED} confirming the absence of discontinuity of $E_d$.  Note that if the energy is discontinuous, its histogram should show a double-peak structure, not a gaussian one.

\begin{figure}
\centerline{\epsfig{file=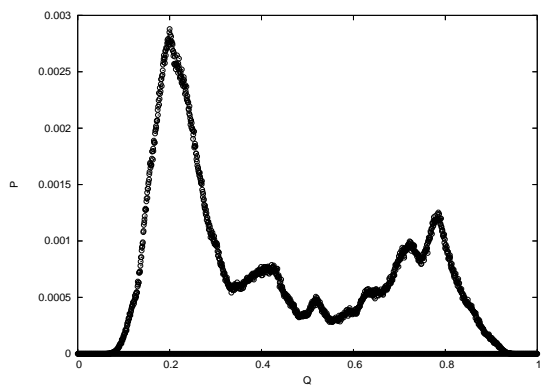,width=2.in,angle=-90}}
\caption{Histogram of the order parameter $Q_d$  taken at $T_c$, for $q=4$, $J_d = 1.6$
with $N_x=1920$, $N_y=60$ and $J =1$. Note the double-peak structure indicating that $Q_d$ is discontinuous at $T_c$. See text for comments.} \label{fig:HQD}
\end{figure}

We have also established  a histogram for $Q_d$ in the following manner.  We divided the interval between 0 and 1 into $N_x$ intervals. At each MC sweep, we added 1 in the interval corresponding to the value of $Q_d$.  In doing so for 2 millions MC sweeps, we obtained a histogram for $Q_d$ which is shown in Fig. \ref{fig:HQD} at $T_c$.
As seen, we have a double-peak distribution of $Q_d$, indicating that during the time the defect line can have both the ordered and disordered phases.  This is a strong signature of the discontinuity of $Q_d$.

Note that the size effects for $N_y=40$, 60, 80 and 100 are not
significant and are included in the error estimation. Simulations
have been carried out also for other values of $J_d$ between 1.2 and 4.  The results
show the same aspects as those shown above with $J_d=1.6$.

\begin{figure}
\centerline{\epsfig{file=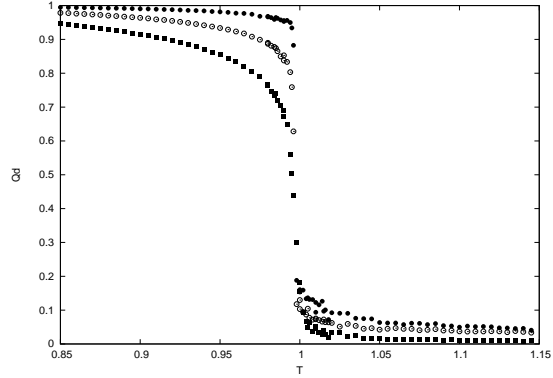,width=2.in,angle=-90}}
\caption{Order parameters  $Q_d$ of the defect line for $J_d = 2$ (upper curve with black circles), $J_d = 1.4$ (middle curve with void circles) and bulk order parameter $Q$ (lower curve with black squares)
vs temperature $T$ for $q=3$
with $N_x=1920$, $N_y=60$ and $J =1$. Note that $Q_d$ is discontinuous at $T_c$. See text for comments.} \label{fig:Q3}
\end{figure}

Now, let us examine the case where $q=3$.  The results are very similar to the case $q=4$ shown above.  So the conjecture of the RG analysis where the transition of the defect line is of first order for $J_q>J$ at $T_c$ for $q>3$ is verified here. Figure \ref{fig:Q3} shows the defect order parameter $Q_d$ versus $T$ for $q=3$ with several values of $J_d$.  Note that $T_c=0.9949$ for $q=3$.

\begin{figure}
\centerline{\epsfig{file=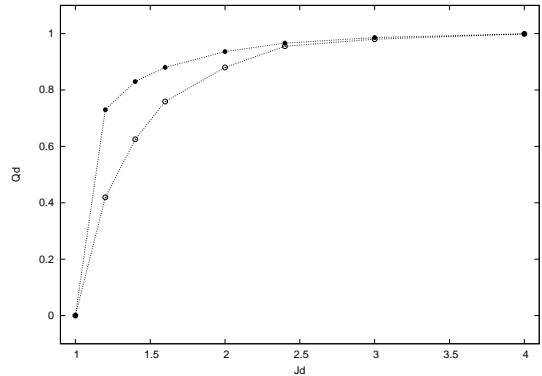,width=2.in,angle=-90}}
\caption{Value of the order parameter $Q_d$  taken at $T_c$ versus $J_d$ for $q=4$ (upper curve) and $q=3$ (lower curve)
with $N_x=1920$, $N_y=60$ and $J =1$. Lines are guides to the eye.} \label{fig:QdJd}
\end{figure}

Let us show in Fig. \ref{fig:QdJd} the value of $Q_d$ taken at $T_c$ as a function of $J_d$. As seen, the gap is increased with increasing $J_d$.

At this stage, it is worth to mention that due to the usual finite size effect, the order parameters $Q$ and $Q_d$ do not vanish above $T_c$: a finite tail exists and decreases with increasing lattice size.  For $Q$, the transition point is, in simulations, taken at the change of curvature of $Q$, i.e. at the maximum of the corresponding susceptibility.  As for $Q_d$, due to its discontinuity at $T_c$, the values shown in Fig. \ref{fig:QdJd} are the upper one at $T_c$. For clarity, the error bars are not shown there, but it is on the second digit, for instance $Q_d=0.880\pm 0.020$ for $J_d=1.6$.

\begin{figure}
\centerline{\epsfig{file=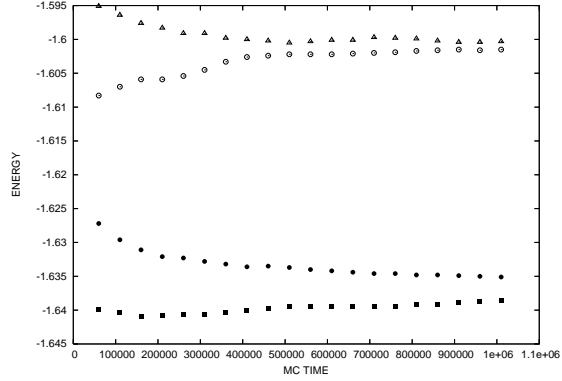,width=2.in,angle=-90}}
\caption{Time dependence of the energy par spin of the defect line for $q=3$ with $J_d = 2$ at $T_c=0.9949$ and at $T=1$. From the bottom: heating to $T_c$, cooling to $T_c$, heating to $T=1$, cooling to $T=1$.
See text for comments.} \label{fig:E-time}
\end{figure}

\begin{figure}
\centerline{\epsfig{file=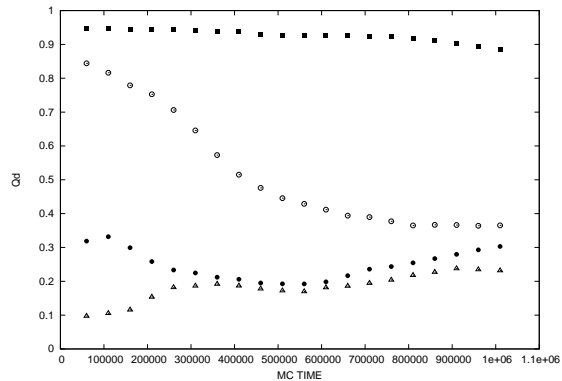,width=2.in,angle=-90}}
\caption{Time dependence of the order parameters  $Q_d$ of the defect line for $q=3$ with $J_d = 2$ at $T_c=0.9949$ and at $T=1$. From the top: heating to $T_c$, heating to $T=1$, cooling to $T_c$, cooling to $T=1$. See text for comments.} \label{fig:Qd-time}
\end{figure}

We show now the time dependence of the order parameter $Q_d$ and the energy.  Figures \ref{fig:E-time} and \ref{fig:Qd-time} show these
quantities  for $J_d=2$ at $T_c=0.9949$ and at a temperature slightly above $T_c$, namely $T=1.$   Two procedures have been used: i) heating, i.e. using the ordered phase (F) as initial spin state ii) cooling, i.e. using a paramagnetic state (P) as initial condition.  We discuss first the energy case.  At $T_c$, heating from the F state (bottom curve in Fig. \ref{fig:E-time}) and cooling from the P state (second curve from the bottom in Fig. \ref{fig:E-time}) give the same energy only after two millions MC steps/spin (see the first million MC steps in Fig. \ref{fig:E-time}).  At $T=1$, one needs almost the same MC time to get the same energy for the heating and cooling (curves three and four from the bottom in Fig. \ref{fig:E-time}).

The time dependence of $Q_d$ is interesting (Fig. \ref{fig:Qd-time}):

i) At $T_c$, $Q_d$ with F condition stays stable (first curve from top in Fig. \ref{fig:Qd-time})  and reaches the stationary value 0.88 at two millions MC steps/spin.  This means that the defect line is ordered at $T_c$ while the bulk spins are disordered. However, when cooled from the P state (third curve from the top), $Q_d$ stays very small (disordered state).  We conclude that at $T_c$ there are two possible values of $Q_d$ for the same energy. This explains the gap of $Q_d$ at $T_c$ shown earlier.

ii) At $T=1$, slightly above $T_c$, $Q_d$ takes a long time ($\simeq 500000$ MC step/spin) to become disordered in the heating procedure (second curve from the top) while it is disordered all the way in the cooling procedure.

Let us discuss more about the discontinuity of the order parameter $Q_d$ at $T_c$. In a standard first-order transition, the ordered and disordered phases coexist at $T_c$  yielding a discontinuity in both energy and order parameter. In the case studied here, only $Q_d$ is discontinuous, while $E_d$ is not.  Thus, there is no double-peak energy distribution.  To answer the question how $Q_d$ can take two values at $T_c$, we have examined the snapshots of the defect line taken during the simulation time.  Interesting enough, we observed configurations schematically of the types:

i) 111111..1111111..111111.. where dots indicate  Potts values other than 1 (same kind of ordered segments separated by disordered portions)

ii) 111111..3333333..222222.. (random ordered segments separated by disordered portions)

It is obvious that these configurations give the same energy if the number of disordered portions is the same, while $Q_d$ is different: the first configuration yields a large value of $Q_d$ and the second a zero one.  This observation explains why at $T_c$ the energy is continuous but the order parameter is not.

The phase transition observed here for the defect line is very interesting in several aspects:

i) the nature of the transition is novel in the sense that only the defect order parameter is discontinuous, not the defect energy, nor the bulk order parameter and bulk energy;

ii) note that long-range interaction in one dimension can cause a first-order transition,\cite{Bayong1999,Reynal2005,Reynal2004} while systems with short-range interaction do not show such a phase transition.  The present defect line with NN  interaction shows thus an exception.  We believe  that its immersion in a disordered neighboring lines at $T_c$ plays a key role in provoking such a pseudo-discontinuous transition.

\section{ Conclusions}

In this paper we have studied a model of a two-dimensional solid with an extended defect line using the Migdal-Kadanoff renormalization group.  Since the elastic energy turns out to be an irrelevant field, we then studied this model for zero elastic energy, i.e. $2d$ Potts model with a defect line, using Monte Carlo simulations.  The renormalization-group analysis of the Potts model on a 2$d$ hierarchical lattice with a defect line, suggests an interesting behavior at the bulk transition $T_c$ if the interaction on the defect line $J_d$ is larger  than the bulk interaction $J$: the order parameter of the defect line should undergo a discontinuity at $T_c$ while the energy is continuous.    Our Monte Carlo simulations of the 3- and 4-state Potts model on the square lattice with a defect line confirm the RG prediction.

To conclude, we would like to emphasize two points of interest. First, the existence of a phase transition in a one dimensional system (defect line) is rare.  It is induced by the bulk infinite correlations at criticality. Second, the nature of the phase transition is unusual in the sense that the order parameter is discontinuous while the energy is not.  This is somewhat similar to the Thouless transition in $1d$ models with long-range interactions.

M.K. wishes to thank the University of Cergy-Pontoise for hospitality while this work was carried out.

{}


\begin{thebibliography}{}

\bibitem{Gomez} L. Gomez, A. Dobry, C. Geuting, H. T. Diep, L. Burakovsky, Phys. Rev. Lett. {\bf 90}, 095701 (2003).

\bibitem{Man} Z. Y. Man, Y. W. Zhang, D. J. Srolovitz, Computational Materials Science {\bf 44}, 86 (2008).

\bibitem{Blumberg} R. L. Blumberg Selinger, Z. G. Wang, W. M. Gelbart, A.
Ben-Shaul, Phys. Rev. A {\bf 43}, 4396 (1991).

\bibitem{Kaufman96} M. Kaufman and J. Ferrante, NASA Tech. Memo. 107112 (1996).

\bibitem{Ferrante} J. Ferrante and J. R. Smith, Phys. Rev. B {\bf 31}, 3427 (1985).

\bibitem{Kaufman2008}  M. Kaufman and H. T. Diep, J. Phys.: Condens. Matter. {\bf 20}, 075222 (2008).

\bibitem{Hassold} G. N. Hassold and D. J. Srolovitz, Phys. Rev. B {\bf 39}, 9273
(1989).

\bibitem{Potts} R. B. Potts, Proc. Camb. Phil. Soc. {\bf 48}, 106 (1952).

\bibitem{Wu} F. Y. Wu, Rev. Mod. Phys. {\bf 54}, 235-268 (1982).

\bibitem{Kaufman84b}M. Kaufman, D. Andelman, Phys. Rev. B {\bf 29}, 4010-4016 (1984).

\bibitem{Kaufman84c}  M. Kaufman and M. Kardar, Phys. Rev. B {\bf 29}, 5053 (1984).

\bibitem{Ivanskoi} V. A. Ivanskoi, Russian Technical Physics {\bf 53}, 455 (2008).

\bibitem{Berker} A. N. Berker and S. Ostlund, J. Phys. C {\bf 12}, 4961-4975 (1979).

\bibitem{Griffiths82} R. B. Griffiths and M. Kaufman, Phys. Rev. B {\bf 26}, 5022 (1982).

\bibitem{Kaufman84d}  M. Kaufman and K. K. Mon, Phys. Rev. B {\bf 29}, 1451 (1984).

\bibitem{Kaufman84} M. Kaufman, R.B. Griffiths, Phys. Rev. B {\bf 30}, 244 (1984).

\bibitem{Erbas} A. Erbas, A. Tuncer, B. Yucesoy, A. N. Berker, Phys Rev E {\bf 72},
026129 (2005).

\bibitem{Guven} C. Güven, A. N. Berker, M. Hinczewski, H. Nishimori,
Phys. Rev. E {\bf 77}, 061110 (2008).

\bibitem{Kaufman82} M. Kaufman and R.B. Griffiths, Phys. Rev. B {\bf 26}, 5282 (1982).

\bibitem{Thouless} D. J. Thouless, Phys. Rev. {\bf 187}, 187 (1969).

\bibitem{Aizenman} M. Aizenman, J. T. Chayes, L. Chayes, and C. M. Newman, J. Stat. Phys. {\bf 50} 1 (1988).

\bibitem{Beale} P. D. Beale and D. J. Srolovitz, Phys. Rev. B {\bf 37}, 5500 (1988).

\bibitem{Fortuin} C. M. Fortuin, P. W. Kasteleyn, J. Phys. Soc. Jpn. Supplm. {\bf 26}, 11
(1969).

\bibitem{Migdal} A. A. Migdal, JETP (SovPhys){\bf 42}, 743 (1976).

\bibitem{Kadanoff} L. P. Kadanoff, Ann. Phys.(NY) {\bf 100}, 359 (1976).

 \bibitem{Berezinskii} V. L. Berezinskii, Sov. Phys. JETP {\bf 32}, 493 (1971).

\bibitem{Kosterlitz}  J. M. Kosterlitz and D. J. Thouless, J. Phys. C {\bf 6}, 1181(1973).

\bibitem{Hinczewski} M. Hinczewski and A. N. Berker, Phys. Rev. E {\bf 73}, 066126 (2006).

\bibitem{Binder} K. Binder and D. W. Heermann, {\it
Monte Carlo Simulation in Statistical Physics}, Springer (2002).

\bibitem{Ferren} A. M. Ferrenberg and R. H. Swendsen, Phys. Rev. Lett.
{\bf{61}}, 2635(1988); Phys. Rev. B {\bf{44}}, 5081(1991).

\bibitem{Bergmann} D. J. Bergmann and B. I. Halperin, Phys. Rev. B {\bf 13}, 2145 (1976) and references therein.

\bibitem{Fisher68}  M. E. Fisher, Phys. Rev. {\bf 176}, 257 (1968).

\bibitem{Diep}  E. H. Boubcheur and H. T. Diep, J. Appl. Phys. {\bf 85}, 6085 (1999);
      E. H. Boubcheur , P. Massimino, H. T. Diep, J. of Magn. and Magn. Mater. {\bf 223}, 163-168 (2001) and references therein.

\bibitem{Landau}  Xiaoliang Zhu, D. P. Landau, and N. S. Branco,
Phys. Rev. B {\bf 73}, 064115 (2006).


\bibitem{Bayong1999}E. Bayong, H. T. Diep, and Viktor Dotsenko, Phys. Rev. Lett. {\bf 83}, 14 (1999); Phys. Rev. Lett. {\bf 85}, 5256 (2000).

\bibitem{Reynal2005} S. Reynal and H. T. Diep, Phys. Rev. E {\bf 72}, 056710 (2005).

\bibitem{Reynal2004} S. Reynal and H. T. Diep, Phys. Rev. E {\bf 69}, 026109 (2004).

\end{thebibliography}
\end{document}